\documentclass[12pt,a4paper]{article}
\usepackage{epsfig}
\usepackage{graphicx}                  
\usepackage{ae}
\usepackage{amsmath}
\usepackage{amssymb}
\usepackage{graphics}
\usepackage{dcolumn}
\usepackage{bm}
\begin{document}
\begin{center}

{\bf \Large Ethernet based Data logger for Gaseous Detectors}\\
\vspace{0.1cm}
S.~Swain$^*$, P.~K.~Sahu and S.~Sahu\\

Institute of Physics, HBNI, Sachivalaya Marg, P.O: Sainik School \\ Bhubaneswar - 751 005, Odisha, India\\
$^*$E-mail: sagarikaswain.swain2@gmail.com
\end{center}

\abstract{A data logger is designed to monitor and record ambient parameters such as temperature, pressure and relative humidity along with gas flow rate as a function of time. These parameters are required for understanding the characteristics of gas-filled detectors such as Gas Electron Multiplier (GEM) and Multi-Wire Proportional Counter (MWPC). The data logger has different microcontrollers and has been interfaced to an ethernet port with a local LCD unit for displaying all measured parameters. In this article, the explanation of the data logger design, hardware, and software description of the master microcontroller and the DAQ system along with LabVIEW interface client program have been presented. We have implemented this device with GEM detector and displayed few preliminary results as a function of above parameters.}

\section{Introduction}

The Gas Electron Multiplier (GEM) is one of the advanced micropattern detectors that plays a significant role in many scientific research areas. The features such as high rate handling capability, excellent spatial and temporal resolutions, stable gain, lower ion backflow fraction and flexibility in designing, make it a better option for researchers \cite{sauli}. Studies of GEM detectors have been initiated, with the main goal of measuring gas gain and optimize its stability over a period of time with total
charge accumulation. The gain of the detector depends on temperature and pressure (T/P) with a unique exponential function \cite{gain}, the detail explanation is given in section 5. Therefore, it is necessary to monitor ambient parameters like temperature, pressure and relative humidity (RH) for the calculation of corrected gain during the experiment \cite{datalog}. Since the detector works in continuous gas flow, monitoring gas flow rate is also relevant. In this report, we have built a compact data logging device that can monitor all these parameters as a function of time. These ambient parameters along with flow rate can be stored in separate files with the timestamp by interfacing through LabVIEW software \cite{NI}.

We would like to monitor the very precise value of temperature, pressure, humidity, and gas flow rate with corresponding time information. Therefore, a microcontroller based high-level electronic circuit designing program is selected. The sensors used in this work give very precise measurements, compatible with digital output data as compared to other commercial devices. The dynamic digital output makes it easy for interfacing with LabVIEW software \cite{NI}. The most important aspect is that the data recorded from the experimental setup need to be with the timestamp for further analysis, the individual devices may not have the same timestamp for a particular set of data. Therefore, we have compiled the individual sensors in such a way that correlated sets of data can be measured and stored for a long time without any interruption. One more advantage is that the whole setup is not system dependent and it can be monitored from any place over the LAN.

The hardware based data acquisition system (DAQ) of the data logger is designed with a microcontroller having a 16$\times$4 line Alphanumeric LCD display unit. The display unit can update within a minimum of 5 seconds and can be made longer up to few minutes. The DAQ is built with an ethernet communication port so that it can remotely access any PC over Wi-Fi or LAN connection.

This paper presents the detail design techniques with circuit diagrams for each component, the function of each component along with the interfacing technique to LabVIEW software. Data obtained with this system coupled to a GEM detector is also presented.

\section{Design Principle}\label{construct}

The instrument is designed with a modular technology. There are one master microcontroller and several slave microcontrollers that are connected to it for a modular operation. The master microcontroller is an advanced ARM Cortex M3 controller operating with 84MHz frequency external crystal \cite{arduino}. It is connected to a hard wired TCP/IP embedded ethernet controller using serial peripheral interface bus (SPI). The temperature sensor (DS18B20), the pressure sensor (BMP180) and humidity sensor (DHT11) having inbuilt ADCs are smart sensors and provide a pre-calibrated digital output. The flow sensor (AWM2100V) is analog with a differential output proportional to the gas
flow.

\begin{figure}[htbp]
	\centering 
	\includegraphics[width=.8\textwidth]{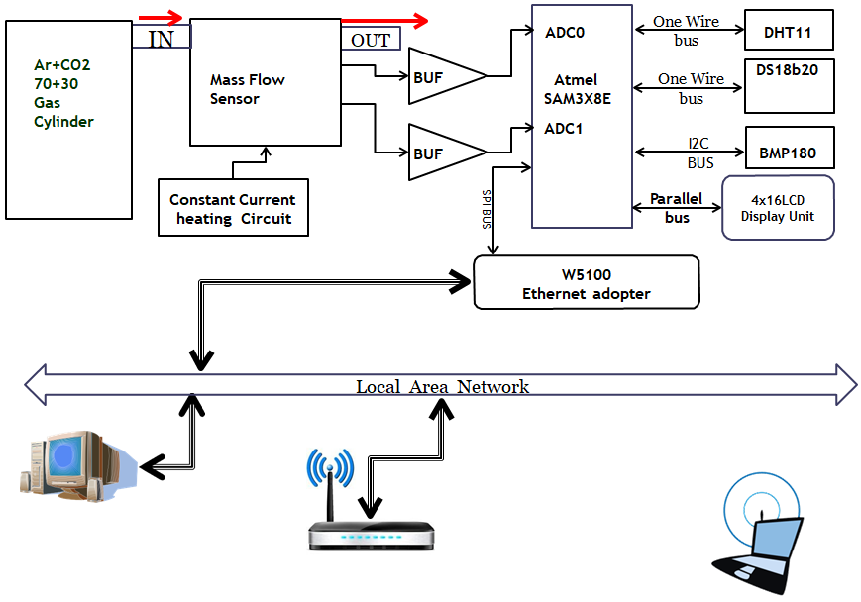}
	\caption{\label{fig:1}Block diagram for the whole setup. All the sensors are connected to the master micro controller with specific communication mode. The ethernet controller is programmed to access the information from the master controller.}
\end{figure}

All these sensors are made accessible by specific communication to the master microcontroller. At first, the humidity and temperature sensors are connected with one wire bus communication mode, where for the pressure sensor, I2C bus communication is used. For the flow rate, the differential output of the flow sensor is connected directly to the master microcontroller using two channels inbuilt ADC. These ADCs have 12-bit resolution with 3.3V reference voltage. The master microcontroller communicates with IP enabled embedded ethernet controller (W5100) and programmed to host an embedded server with a unique MAC, IP, subnet and gateway addresses. Then, it collects all the information from pressure, humidity and temperature sensors using different communication protocols. The block diagram for the whole setup with different interfaces is given in Figure \ref{fig:1}.

All these sensors are pre-calibrated and provide digital data to the master microcontroller. The flow sensor is initially calibrated only for pure gases. Since we are using mixed gas, we need to recalibrate the flow sensor for our experimental purposes. Therefore, we have done a manual calibration by water displacement method for the gas mixture Ar/CO$_2$ in 70/30 ratio. The flow sensor provides a differential voltage, which is then fed to the controller using two 12 bit ADCs.

The controller reads both ADC1 and ADC2 and stores the differential values in an array. Measured data (voltage) is accepted if it falls within a determined range limit defined by the user. The output voltage is converted to the corresponding flow rate by using the fitting formula from the calibration plot discussed in section 3.3.4. There is a 4x16 local LCD unit, where all the ambient parameters, i.e., the temperature in $^{\circ}\mathrm{C}$, atmospheric pressure in mBar, the relative humidity in \% and gas flow rate in SCCM (Standard Cubic Centimeter per Minute) are displayed. The master microcontroller continues to update the LCD display in a certain interval specified by the user through LabVIEW software interface.

In the frontend, LabVIEW based client program is written to request the embedded web server for the information from the sensors. The web server fetches all these informations and writes a long string to an HTTP port. The LabVIEW based program then extracts these informations and displays them in numeric and graphical modes in the computer screen. Also, it is programmed to write temperature in  $^{\circ}\mathrm{C}$, atmospheric pressure in mBar, the relative humidity in $\%$ and gas flow in SCCM unit in a file with fixed time interval as defined by the user.

\section{Hardware Description}

The data logger is configured with three main hardware components:  master microcontrollers, embedded ethernet controller, and sensors. Their design and detailed descriptions are discussed hereafter.

\subsection{Master micro controller}

The master microcontroller is designed with a microcontroller board based on Atmel SAM3X8E ARM Cortex-M3 CPU \cite{micro}. It is a 32 bit ARM core microcontroller, inbuilt USB OTG capability and operating with 84MHz frequency. It has 12 bit ADC with 3.3V reference voltage resulting in an ADC resolution of 0.8mV approximately. Also, it has 512kB flash memory and 98kB SRAM in two separate blocks. The Cortex M3 processor is built on a high-performance processor core, 3 stages pipeline Harvard architecture, which makes it ideal for demanding embedded applications. The processor delivers exceptional power efficiency through an efficient instruction set and extensively optimized design, which provides a high-end processing hardware including single cycle 32x32 multiplications and dedicated hardware division. The architecture for the master microcontroller is same as figure 1 in Ref. \cite{arduino pic}.

\subsection{Embedded ethernet controller}

\begin{figure}[htbp]
	\centering 
	\includegraphics[width=.5\textwidth]{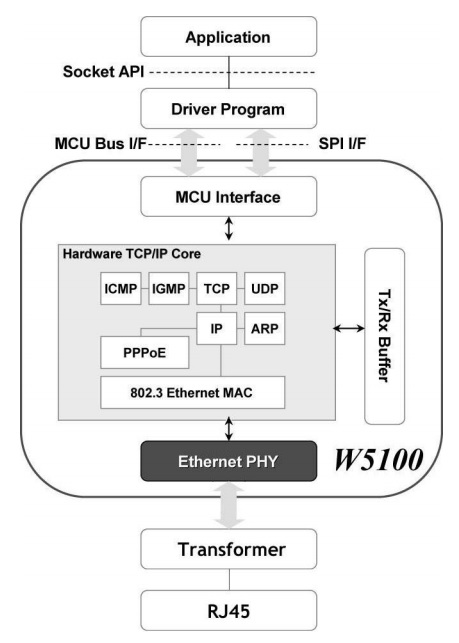}
	\caption{\label{fig:2} 
		Block diagram for the W5100 chip used as a Hardwired TCP/IP embedded ethernet controller.}
\end{figure}

The embedded ethernet controller (W5100) \cite{w5100,w5100.1} is a full-featured single chip internet enabled 10/100 ethernet controller designed for embedded applications,
where it is easy to integrate. The detail block diagram of the chip is given in Figure \ref{fig:2}.
Upon restart the embedded ethernet controller, it is initialized by the master controller with specific MAC, IP, subnet and gateway addresses. Here the controller is initialized as a web server. When any client requests the server, it collects the data from the master controller and sends to an HTTP port. Later the server is initialized to HTTP default port.

\subsection{Sensors}

Temperature, pressure, humidity, and flow sensors are used in the system. The description is given below.

\subsubsection{Temperature Sensor}

\begin{figure}[htbp]
	\centering 
	\includegraphics[width=.8\textwidth]{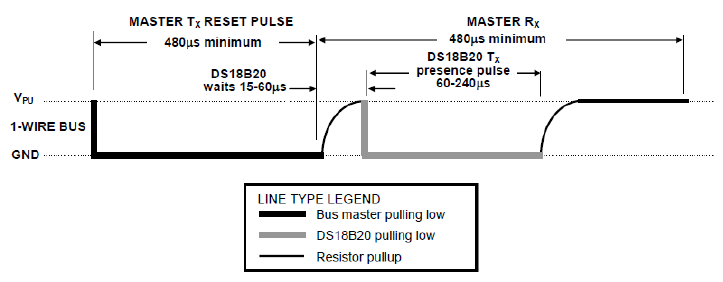}
	\caption{\label{fig:3} 
		Block diagram for the Temperature sensor.}\label{block}
\end{figure}

The temperature sensor (DS18B20) \cite{temp, temp.1} provides 9-bit to 12-bit Celsius temperature measurement and has an alarm function with nonvolatile user programmable, upper and lower trigger points. It communicates over a one-wire bus that requires only one data line for communication with a central microprocessor. This sensor is also pre-calibrated providing enough accuracy for the purpose of the tests. The block diagram is given in Figure \ref{fig:3}. The range is from -55$^{\circ}\mathrm{C}$  to +125$^{\circ}\mathrm{C}$
(-67$^{\circ}\mathrm{F}$ to +257$^{\circ}\mathrm{F}$) and accuracy from -10$^{\circ}\mathrm{C}$ to +85$^{\circ}\mathrm{C}$.

\subsubsection{Pressure Sensor}

\begin{figure}[htbp]
	\centering 
	\includegraphics[width=.5\textwidth]{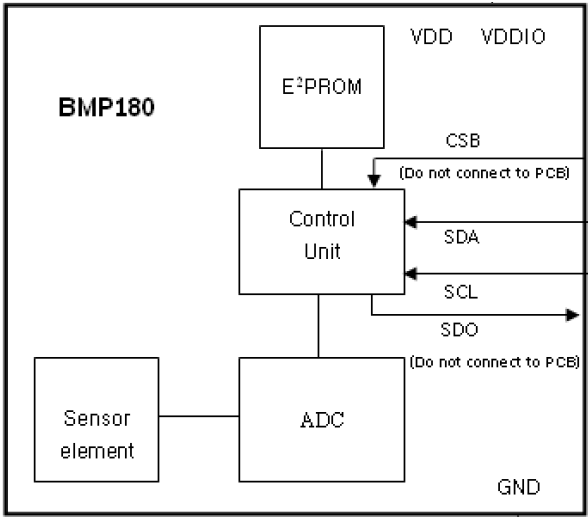}
	\caption{\label{fig:4} 
		Block diagram for presure sensor.}\label{algorithm_MC}
\end{figure}

The pressure sensor (BMP180) \cite{pressure, pressure.1} consists of a piezoresistive sensor, an analog to digital converter, a control unit with E2PROM and a serial I2C interface. The E2PROM has stored 176 bit of individual calibration data, which is used to compensate offset, temperature dependence and other parameters of the sensor. The microcontroller sends a start sequence to a sensor and after converting the time, the resultant value can be read via the I2C interface. The calibration data has been used for calculating pressure in hPa. These constants can be read out from the internal E2PROM via the I2C interface at software initialization.  For dynamic measurement,
the sampling rate is used to increase up to 128 samples per second (standard mode). In this case, it is sufficient to measure the temperature only once per second and to use this value for all pressure measurements during the same period. The block diagram is shown in Figure \ref{fig:4}.

\subsubsection{Humidity Sensor}

\begin{figure}[htbp]
	\centering 
	\includegraphics[width=.8\textwidth]{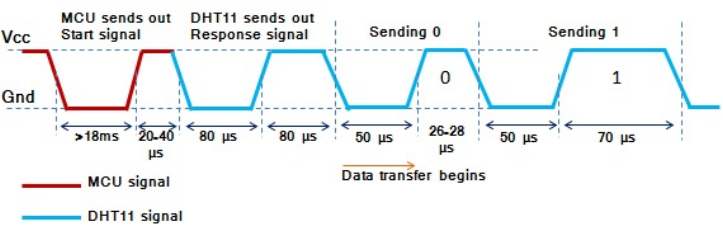}
	\caption{\label{fig:5} 
		Data communication of humidity sensor.}\label{circuit}
\end{figure}

The humidity sensor (DHT11) \cite{humidity} used in this device is a commercially available and a pre-calibrated product. The working principle is as follows. There is a
moisture sensitive substrate between two conductive plates.
The substrate upon exposure to the moisture changes its resistivity. The resultant change is measured by a dedicated microcontroller, which is then communicated to
the master controller using one wire communication mode.
The DHT11 sensor measures relative humidity with repeatability $\pm$1\%  and accuracy 25$^{\circ}\mathrm{C}\pm5\%$. The  data communication of this sensor  is
given in Figure \ref{fig:5}.

\subsubsection{Flow Sensor}

The AWM2100V microbridge mass airflow sensor \cite{flow} is a passive device consists of two Wheatstone bridges. The output voltage from the sensor is proportional to the input gas flow. The voltage range is from 0mV to 44.5mV but not in a linear manner and it can measure up to 200 SCCM. The sensor is externally connected to a constant current circuit and a low ripple power supply for its functionality. The constant current circuit is designed to bias one of the element of the bridge to heat up and keep a constant rate. The bridge is then biased with a low ripple DC voltage to adjust null at the output. The 10V supply source is having a ripple less than 10mV. The constant current heater control circuit and biasing circuit are given below in Figure \ref{fig:6}.

\begin{figure}[htbp]
	\centering 
	\includegraphics[width=.32\textwidth]{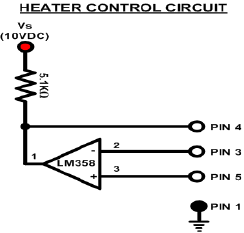}
	\qquad
	\includegraphics[width=.3\textwidth]{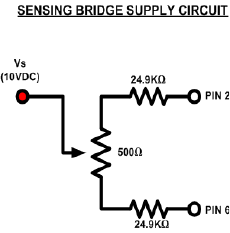} 
	\caption{\label{fig:6} The left panel of figure shows Constant current heater control circuit and 
		in the right panel Biasing circuit is given.}\label{algorithm_LabView}
\end{figure}

\begin{figure}[htbp]
	\centering
	\includegraphics[width=.9\textwidth]{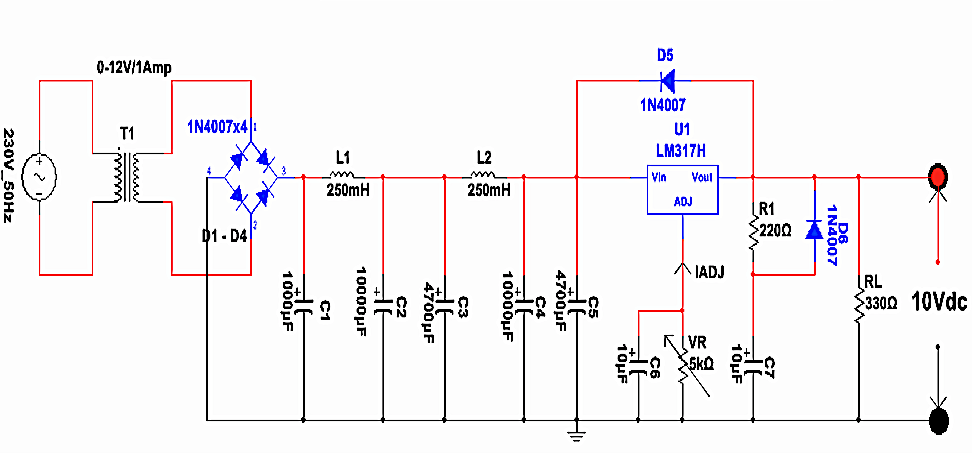}
	\caption{\label{fig:7}
		The low ripple power supply with series LC filter.}\label{algorithm_LabView}
\end{figure}

The low ripple power supply is designed with a series LC filters with a minimum load current of 30mA or more. The power supply scheme has three stages, first, the AC mains is converted into a low DC voltage and a bridge rectifier. The second stage consists of two Pi filters and finally, a voltage regulator is connected to the obtained 10V supply. The PCB designed for this purpose is shown in Figure \ref{fig:7}. This circuit board is connected separately to the flow sensor to make it operational.

The calibration and working principle are described below.

\begin{figure}[htbp]
	\centering 
	\includegraphics[width=.7\textwidth]{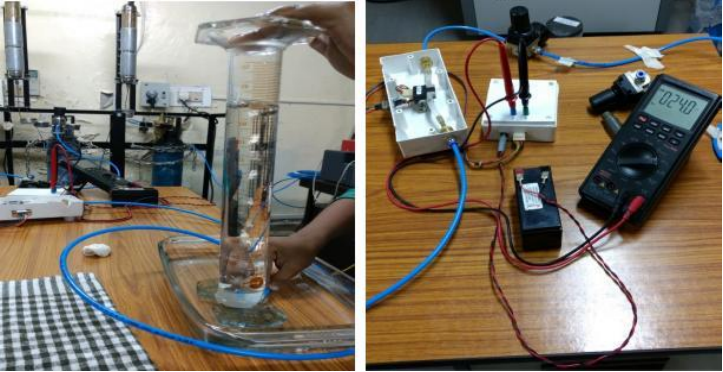}
	\caption{\label{fig:8} 
		Calibration setup for the humidity sensor. In the left panel, the displacement of a fixed volume of the gas inside the cylinder is measured with time; in the right panel the ouput voltage of the sensor is recorded from the multimeter. }\label{algorithm_LabView}
\end{figure}

The output of the gas flow sensor is not linear because the data of flow rate versus differential voltages are for pure gases. Since the test system will use the gas mixture Ar/CO$_2$ (70/30), calibration of the sensor output voltage (in mV) with the mixture flow rate (in SCCM) is needed. The calibration setup is shown in Figure \ref{fig:8} and the method is described hereafter.

At first, a measuring cylinder is filled with water and kept inverted inside a half filled water vessel. One end of the gas outlet tube from the flow sensor is now gently put into the measuring cylinder. At a constant flow rate, the displacement of a fixed volume of the gas inside the cylinder is measured and the time interval is also recorded. The gas flow rate is calculated by the following formula.

\begin{equation}\label{eq:1}
Gas~ flow ~rate = \frac{displaced~ water~ volume}{ measured~ time ~interval}~~ (ml/min)
\end{equation}

\begin{figure}[htbp]
	\centering 
	\includegraphics[width=.9\textwidth]{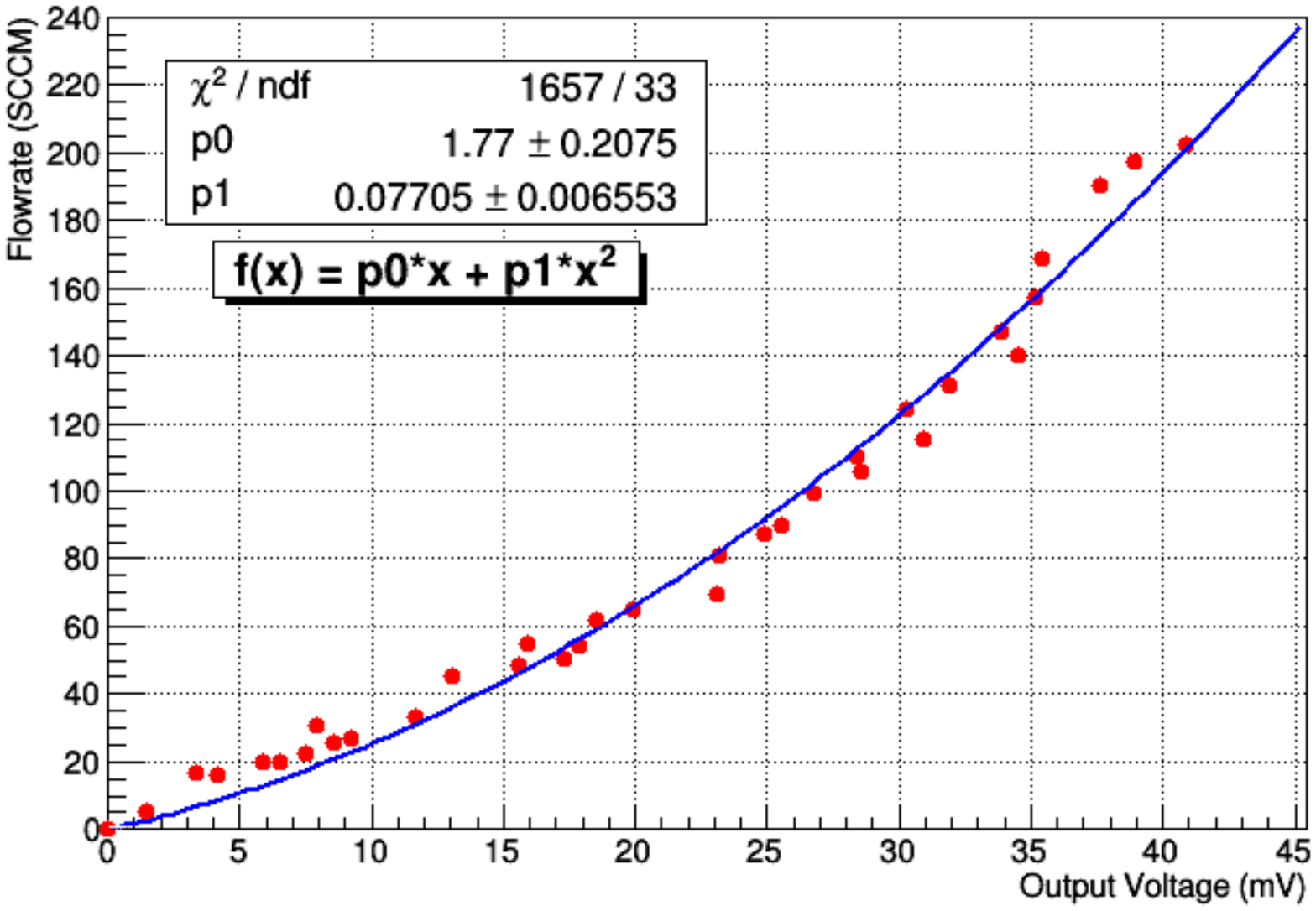}
	\caption{\label{fig:9} 
		The output voltage from the sensor is plotted with corresponding flow rate value observed by water displacement method. The plot is then fitted with a user-defined function to get the value of parameters.}\label{algorithm_LabView}
\end{figure}

The corresponding output voltage (in mV) of the sensor is noted from a multimeter. We repeat this procedure for different gas flow rates up to 200 SCCM. The plot is obtained for the different flow rates as a function of output voltages, which is then fitted with a user-defined function ($ f(x) = po\times x$ + $p1\times x^2$) as shown in Figure \ref{fig:9}. Here, the fitting parameters are used to calculate the corresponding flow rate value from the voltage output of the sensor.

The gas flow sensor works with the principle of heat transfer. The precise design of the flow sensor allows the gas flow over the sensing element so that, the rate of gas flow proportionally cools the sensing elements.  A thermally isolated microstructure bridge is developed with silicon thin film technology. The microbridge with a constant current heating setup inside a precisely designed casing is used as a gas flow sensor and it works with the principle of the mass flow sensor. The arrangement and placement of mass flow sensor are optimized for the fast and sensitive response for any change of gas flow in the sensor. The dual sensing method is used for indication of flow rate and direction of the gas flow inside the sensor.

\begin{figure}[htbp]
	\centering 
	\includegraphics[width=.6\textwidth]{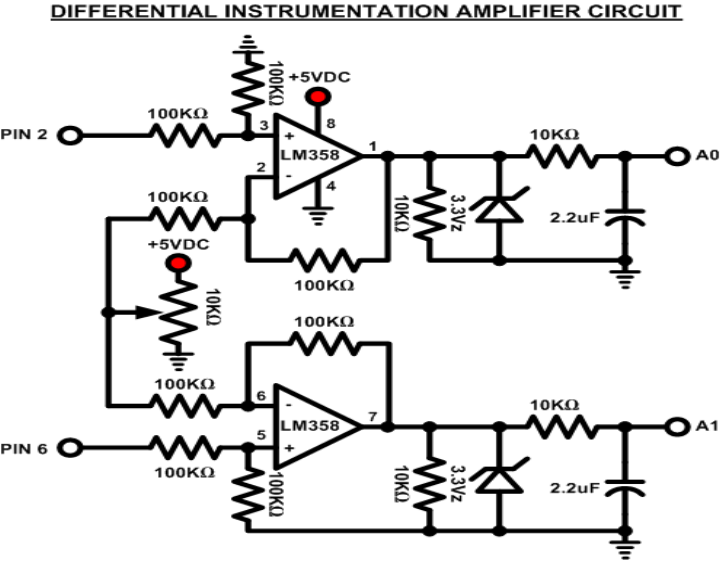}
	\caption{\label{fig:10} 
		The differential instrumentation amplifier circuit.}\label{algorithm_LabView}
\end{figure}

The constant current heating setup minimizes the shifting of a null point due to ambient temperature variation. The constant current heating circuit keeps the temperature at a constant differential (160$^{\circ}\mathrm{C}$) above ambient air temperature, which is sensed by a heat sunk resistor on the chip. The flow of gas transfers heat from one of the elements in the bridge and makes the bridge unbalanced and a proportional bridge output is obtained. This ratiometric output voltage is an indication of gas flow magnitude. Another bridge physically oriented in the casing, so that the output voltage polarity will decide the direction
of gas flow. The circuit is given in Figure \ref{fig:10}.

\begin{figure}[htbp]
	\centering 
	\includegraphics[width=1.0\textwidth]{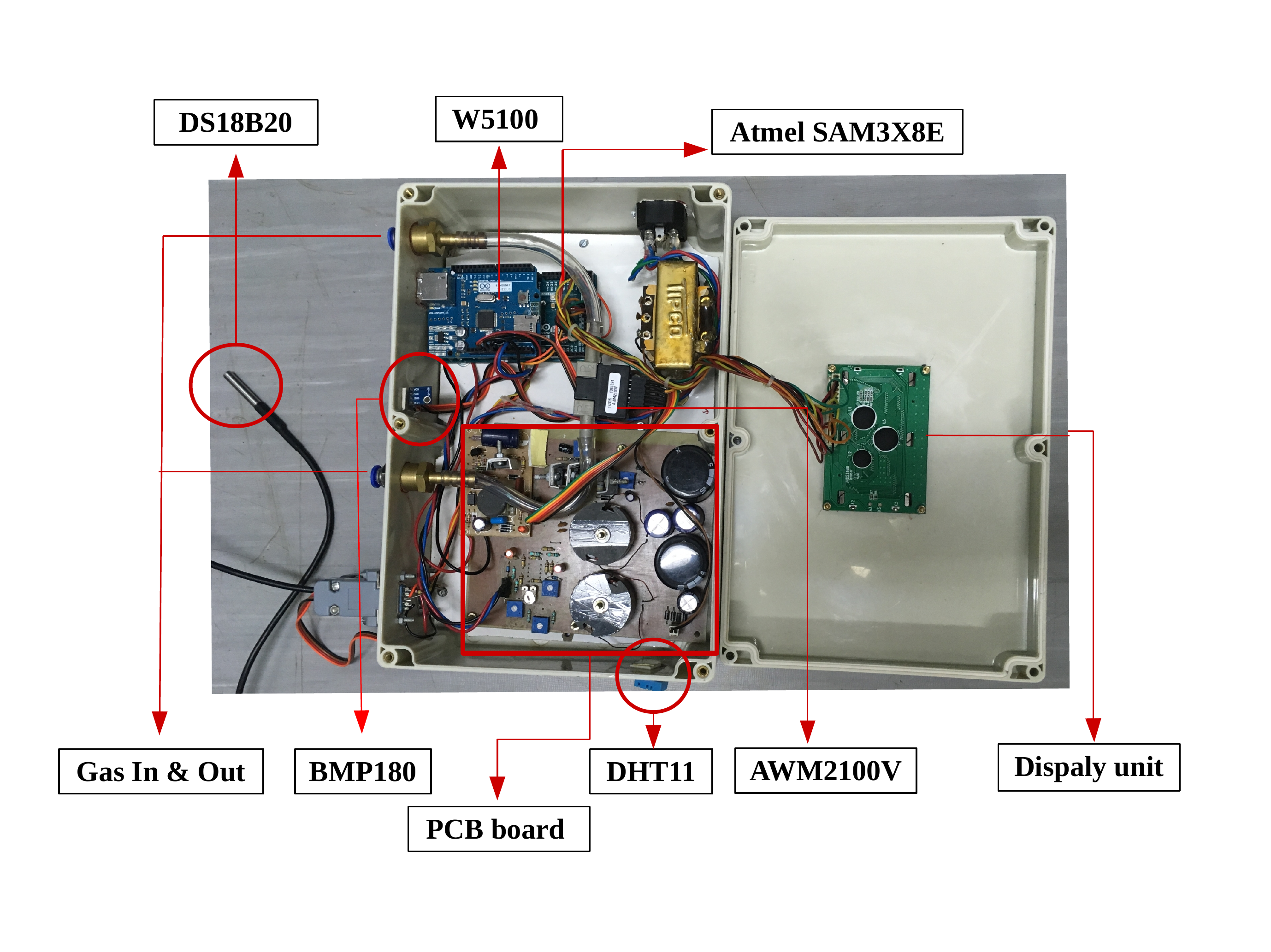}
	\caption{\label{fig:11} The final electronic set up with temperature (DS18B29), pressure (BMP180), humidity (DHT11), flow (AWM2100V) sensors, ethernet adopter (W5100), arduino board (Atmel SAM3X8E), display unit and PCB contaning constant current heating circuit, 10V power supply with less than 10mV ripple and amplifier circuit for flow sensor.}\label{algorithm_LabView}
\end{figure}

The final electronic setup of assembled circuit is given in Figure \ref{fig:11}.

\section{Software Description}

\subsection{Algorithm for master microcontroller}

\begin{figure}[htbp]
	\centering 
	\includegraphics[width=.7\textwidth]{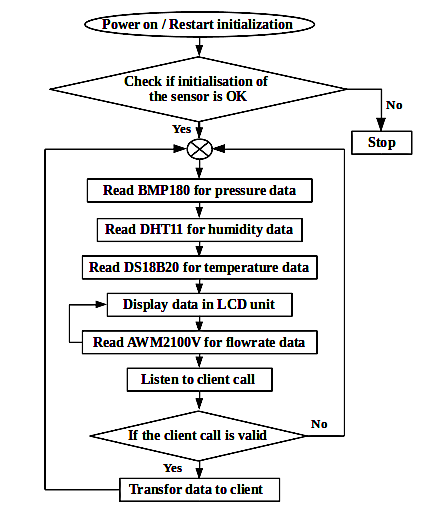}
	\caption{\label{fig:12} 
		Functioning algorithm of the master micro controller.}
\end{figure}

The function of master controller is given in a flowchart in Figure \ref{fig:12}.

\begin{itemize}
	
	\item Power on restart initialisation:  
	
	When the system is switched on, the master controller initialises the pressure by I2C bus, temperature and humidity sensors with one wire bus.

	If initialization fails at any stage, then it will indicate in LCD and stop the process there.
	
	\item Reading the pressure sensor:
	
	The master controller sends a request to the I2C bus to the pressure sensor BMP180. The pressure sensor reads temperature and pressure of the environment and sends a temperature compensated atmospheric pressure value in mBar. The reading of atmospheric pressure is structured as a subroutine which returns a float value to the main program.
	
	\item Reading relative humidity sensor:
	
	After completing the above task, the master controller sends a request over one wire bus to the relative humidity sensor DHT11. The sensor reads and sends the relative humidity value in \%. The reading is also structured as a subroutine, which returns a float value to the main program.
	
	\item Reading of temperature sensor:
	
	The master controller sends a request to the temperature sensor DS18B20. The temperature sensor reads the temperature and sends to master controller via one wire communication port.

	\item Displaying the data:
	
	The controller displays the temperature, pressure and relative humidity in 4 Line LCD display using 4-bit data communication mode.
	
	\item Reading gas flow sensor:
	
	The master controller is configured with ADC0 and ADC1 for 12-bit resolution and the output of gas flow sensor is generating a differential voltage in mV order.  As the output is very low, to filter noise from measurement a statistical method is adopted. The statistical process for filtering the data is as follows: two arrays consisting of 50 data points were built for ADC0 and ADC1. For consistency of the data, the standard deviation of each array is checked. If the standard deviation of the data points for both the array is less then 5 ADC counts, i.e., 4.02mV then the data sets are accepted otherwise rejected and escape from the subroutine with 0 value. When the data points of each array is found to be consistent, the mode of the data array is calculated. That value is converted into a flow rate using a formula given below:
	
	Mode of ADC0 Array=m0
	
	Mode of ADC1 Array=m1
	
	Differential ADC count d = abs(m0-m1)
	
	Differential output in mV x = d(3300/4096)
	
	flow=(1.77x+(0.07705x$^2$))/1.12                     [factors obtained from fitting]
	
	A constant factor 1.12 is divided to nullify the differential amplifier gain factor. This flow value is displayed on the LCD.
	
	\item  Listen to Client call:
	
	The master controller communicates with the embedded ethernet controller for a request of any specific client, if the client requests data from this server, the data string containing all information is transferred. If the client call is invalid or no client call, then the system goes back to the second step to start a new loop.

\end{itemize}

\subsection{Algorithm for DAQ system}

\begin{figure}[htbp]
	\centering
	\includegraphics[width=.9\textwidth]{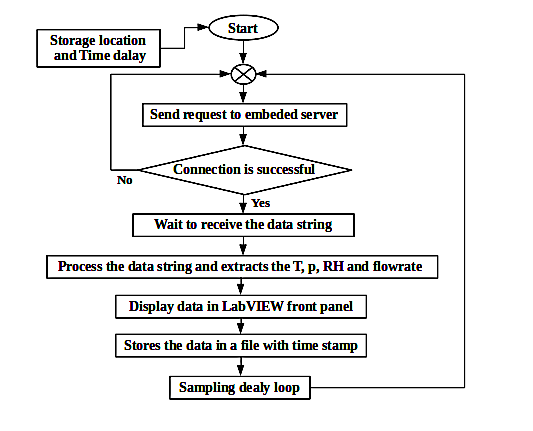}
	\caption{\label{fig:13}
		Algorithm for the DAQ program.}
\end{figure}

Data acquisition is the process of sampling signals that are measured and converting into digital numeric values.  These values can be manipulated in the computer according to users.  Here, the DAQ program is developed with LabVIEW platform and the algorithm is shown in Figure \ref{fig:13}.

\begin{itemize}
	
	\item Program Starting:
	When the program is started, it manually accepts the location for data storage and time delay set for each interval.
	
	\item Requesting the Server:
	The client requests the server for data with a specific format i.e. 10.0.100.38/data, on receiving this, the server sends a string. If the connection is successful it waits to receive the string otherwise back to start a new loop.
	
	\item Processing the String:
	The received string is formatted in HTML and it can be directly displayed in a web browser. So, the entire string is consisting of information of temperature, pressure, relative humidity and gas flow rate along with HTML body. From this string, all the individual data is separated and displayed in the front panel.
	
	\item Storing in File:
	It is programmed to collect the time information from the PC time setup and store in a file indicated in the file browser option.
	
	\item Sampling Interval:
	The time interval manually programmed at the time of initialization and will execute a delay loop. After finishing the delay loop, the control goes to the second step of the DAQ algorithm.
	
\end{itemize}

\subsection{Interfacing with LaBVIEW software}

LabVIEW, which stands for Laboratory Virtual Instrument Engineering Workbench, is a software platform to create
application programs using a graphical notation \cite{NI}.
The LabVIEW programming is different from traditional programming languages like C, C++, or Java, these are basically script based programming. LabVIEW is an industry standard software for integrating hardware and designing applications. It can create programs that run on those platforms, as well as Palm OS and different embedded platforms including Field Programmable Gate Arrays (FPGAs), Digital Signal Processors (DSPs), microprocessors, etc.

\begin{figure}[htbp]
	\centering 
	\includegraphics[width=1.0\textwidth]{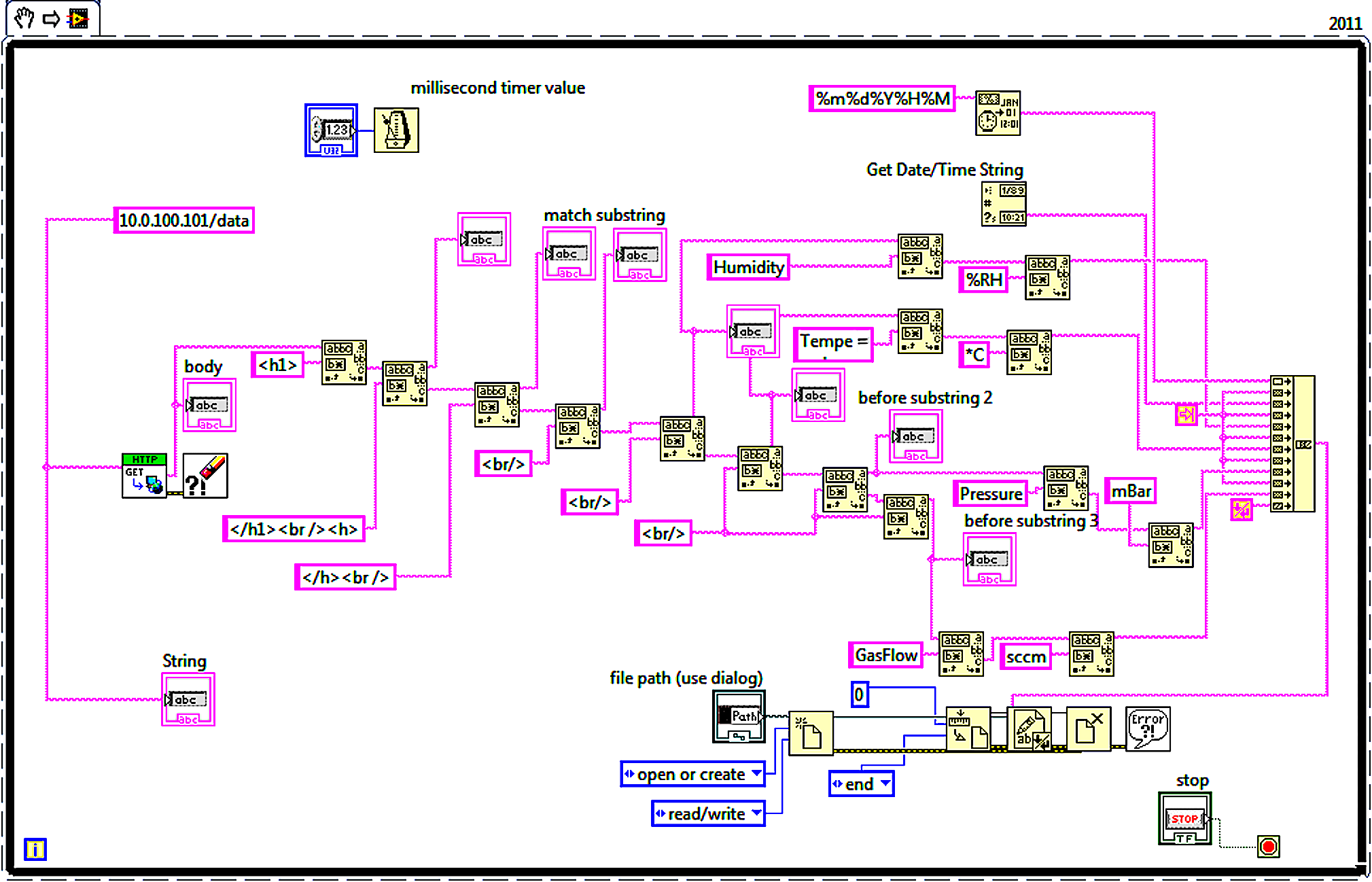}
	\caption{\label{fig:14} 
		Block diagram showing graphical source code.}
\end{figure}

\begin{figure}[htbp]
	\centering 
	\includegraphics[width=.9\textwidth]{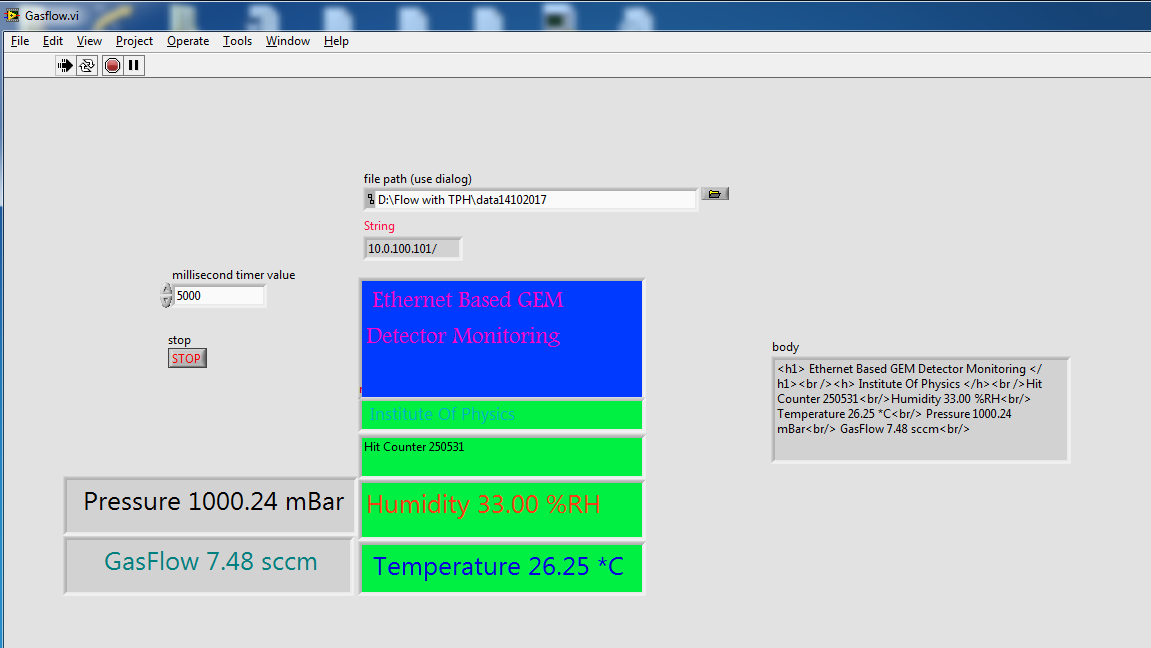}
	\caption{\label{fig:15} 
		Front panel with display units and control input.}
\end{figure}

\begin{figure}[htbp]
	\centering 
	\includegraphics[width=.9\textwidth]{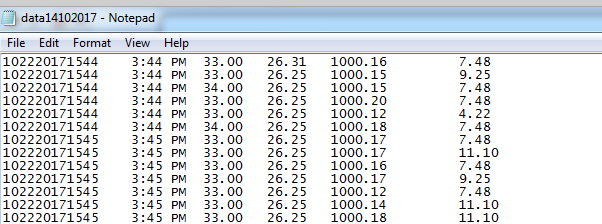}
	\caption{\label{fig:16} 
		Snapshot of the file used for storing the data.}
\end{figure}

The LabVIEW program files have .vi extension that stands for virtual instruments (VIs) and a VI has two main parts: a front panel and a block diagram. These are shown in Figure \ref{fig:14} and Figure \ref{fig:15}, respectively.

The front panel is the interactive user interface of VI, this part appears to the user to display data and provide control input. The controls appear in the front panels and display settings also can be partially changed during the execution of the program.

The block diagram is the VI's source code and it is developed in LabVIEW's graphical programming language. It is the executable program blocks arranged in the data flow sequence. The constituent components of a block diagram are lower level VIs, built-in functions, constants, and program execution control structures.  The block constant, function and structures appearing in block diagram cannot be changed during the execution of the program. The highlighted execution of the program shows the signal, data flow sequence, and intermediate values.

The DAQ software developed in this application is basically an HTTP client VI that interacts with embedded web servers, hosted in the LAN.  As explained before, the embedded web server collects all the data from different sensors and converts it into a string added HTTP headers and body. 
Then it sends a web request by GET HTTP command to return the headers and body data from the server. The header and body data are then separated using match string function. The temperature, pressure, relative humidity and gas flow information are also separated and displayed with a timestamp in a specified file, set in the front panel. The data storing format for our set up is shown in the Figure \ref{fig:16}. The leftmost column is date and time with 24 hours frame and then time with 12 hours frame, the relative humidity in $\% $, the temperature in $^{\circ}\mathrm{C}$, pressure in mbar, and flow rate in SCCM units, respectively.

\begin{figure}[htbp]
	\centering 
	\includegraphics[width=.8\textwidth]{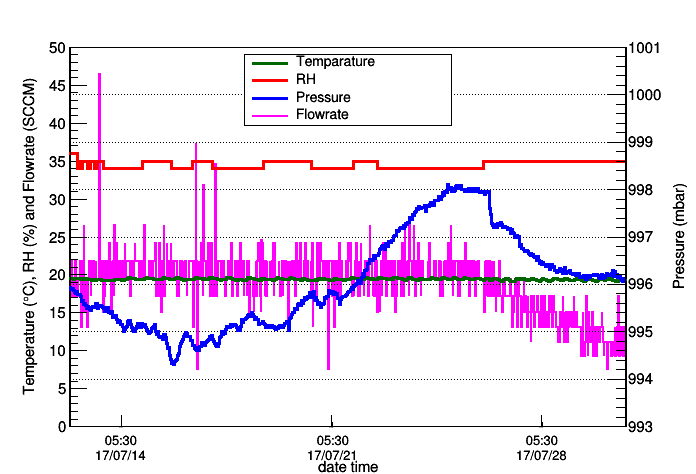}
	\caption{\label{fig:17} 
		Temperature, pressure, RH and flow rate are plotted as a function of date-time.}
\end{figure}

\begin{figure}[htbp]
	\centering 
	\includegraphics[width=.65\textwidth ]{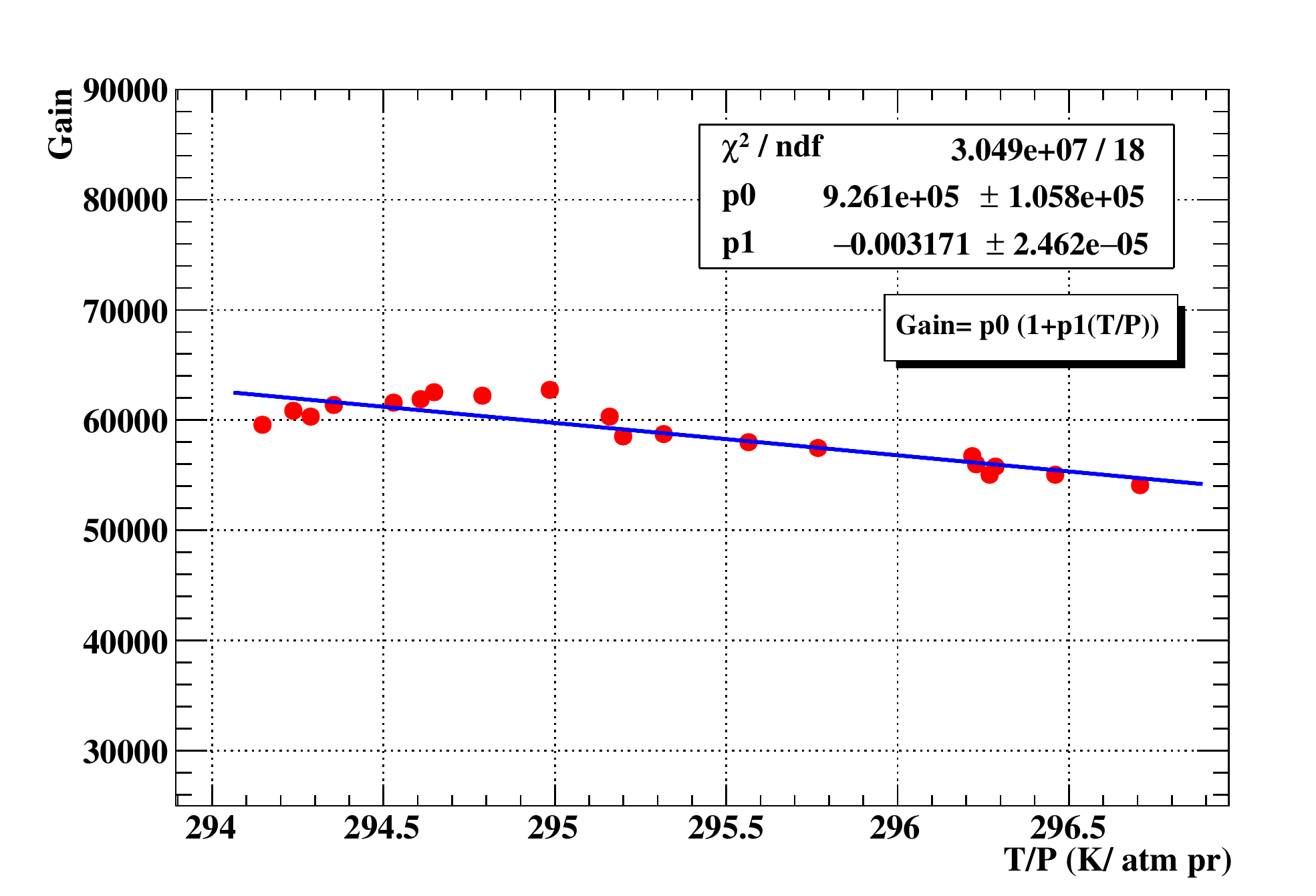}
	\caption{\label{fig:18} 
		Gain of the GEM detector versus T/P.}
\end{figure}

\begin{figure}[htbp]
	\centering 
	\includegraphics[width=.65\textwidth ]{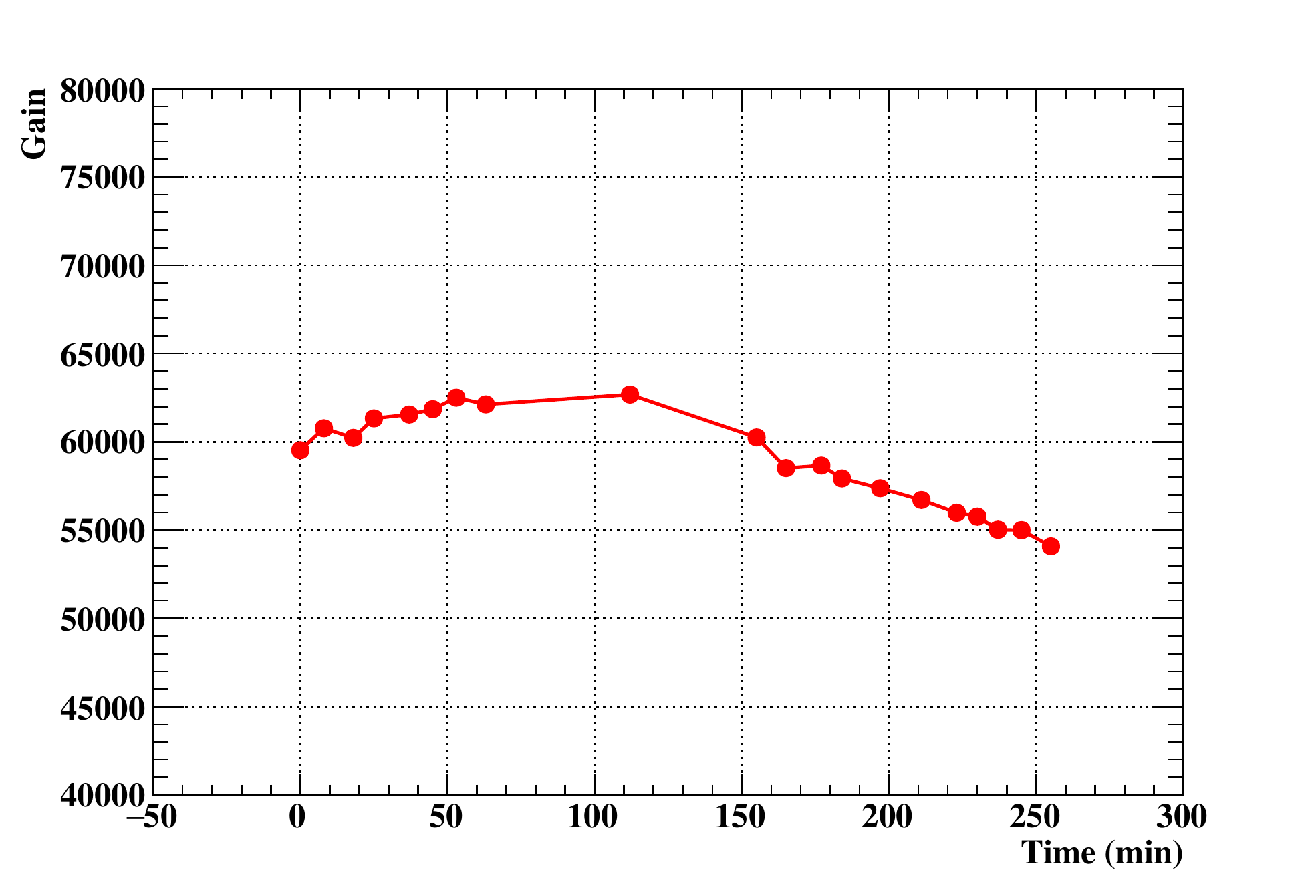}
	\caption{\label{fig:19} 
		Variation of measured gain with time.}
\end{figure}

\begin{figure}[htbp]
	\centering 
	\includegraphics[width=.65\textwidth ]{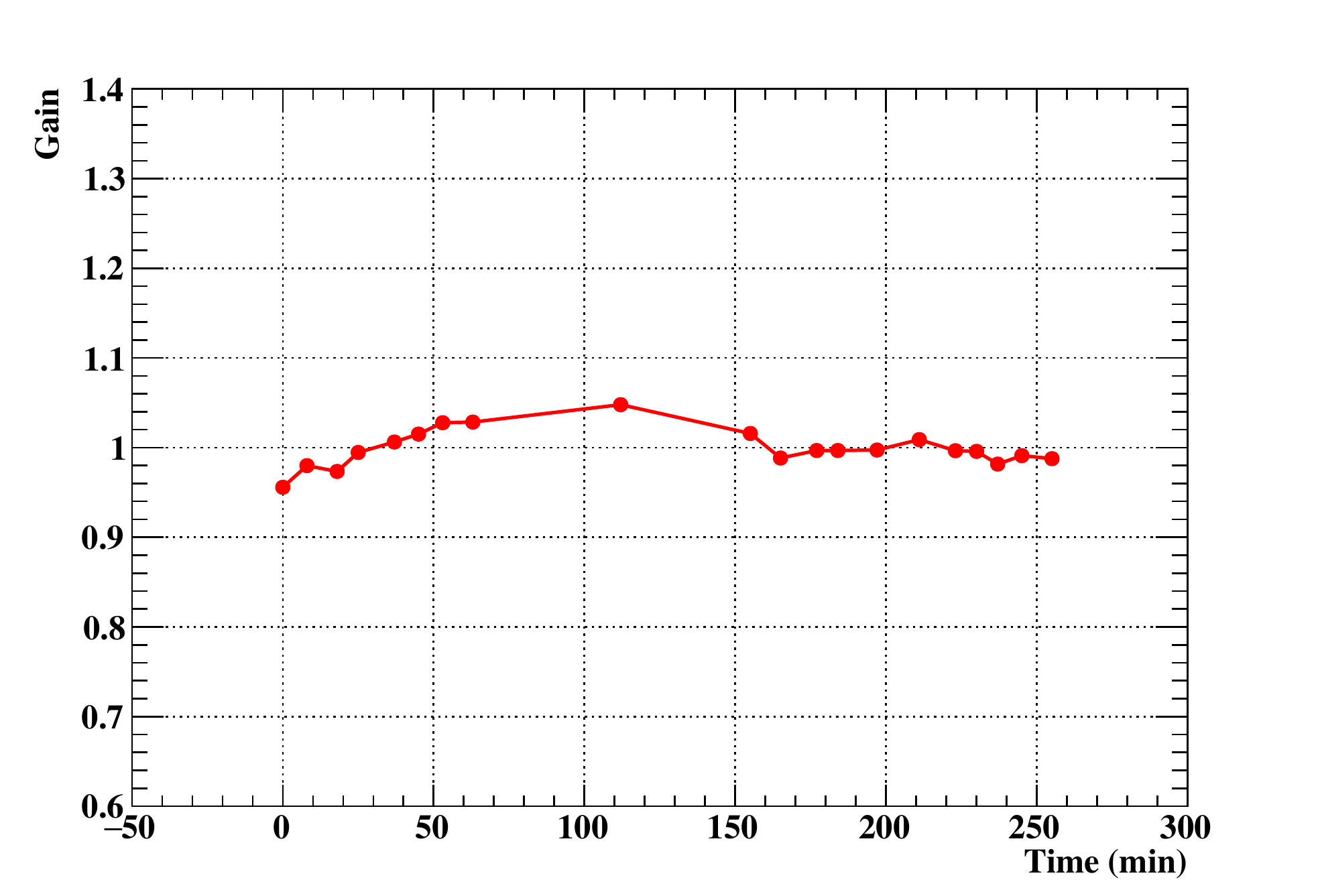}
	\caption{\label{fig:20} 
		Variation of normalised gain with time.}
\end{figure}

\section{Results}

The DAQ setup explained in the previous section has been used to monitor the behaviour of a GEM detector. The gain of the detector depends exponentially on the T/P ratio. Gain can be expressed as a function of first Townsend coefficient, which is the inverse of the mean free path in ionization process.
It can also be associated with the average distance between two consecutive collisions that an electron covers inside the medium. This, in turn, depends on gas temperature (T), pressure (P), relative humidity (RH) and flow rate. The Townsend coefficient is also inversely proportional to the gas density, thus from the ideal gas law gain of the detector changes with absolute temperature T in Kelvin and pressure P in atmospheric pressure as

\begin{equation}\label{eq:2}
G(T/P)={Ae^{B\frac{T}{P}}},
\end{equation}

where A and B are fitting parameters. This is obtained from the fitting of measured gain and $\frac{T}{P}$ by the exponential function \cite{gain}.

The plot of all these ambient parameters with time interval is given in Fig.\ref{fig:17}. Here the data are recorded for the constant voltage applied to a GEM detector prototype.

The horizontal axis shows the date and time whereas the left vertical axis shows the temperature in $^{\circ}\mathrm{C}$ and RH in $\%$ and flow rate in SCCM unit. The right vertical axis shows the atmospheric pressure in mbar. Another important parameter for the GEM detector is gain (Eq.\eqref{eq:2}) and is plotted in Figure \ref{fig:18} as a function of ambient parameters (T/P).

The measured gain and the ratio T/P is also fitted with the exponential function. We have taken the exponential function up to maximum two order terms and the fitting parameters obtained are used to calculate the normalised gain by using the following relation:

\begin{equation}\label{eq:3}
Gain_{normalized}=\frac{Gain_{measured}}{A(1+B(T/P))}.
\end{equation}

The raw gain and the normalized, corrected gas gain
are plotted as a function of time as shown in Figure \ref{fig:19} and Figure \ref{fig:20}. There are some fluctuations in the normalised gain but no steady decrease with time is observed.

\section{Conclusion}
An ethernet based data logger has been developed to monitor the relevant parameters when testing gaseous detectors. It monitors ambient parameters such as temperature, pressure and relative humidity and gas flow rate. Using this data logger, we can continuously store these parameters with a timestamp in a separate file. This system is very user-friendly and cost-effective when used for any gaseous detector. One more advantage is that it can be communicated through an ethernet port that makes it easily accessible to the user. The LabVIEW interface client program allows to easily write graphical source code, control input as well as to develop a comfortable display unit.


\begin{thebibliography}{99}
	
	\bibitem{sauli} F. Sauli, \emph {The gas electron multiplier (GEM): operating prin-
		ciples and applications}, \emph{NIM. A}  {\bf 805} (2016) 2.
	
	\bibitem{gain} M.C. Altunbas et al., \emph{Aging measurements with the Gas Electron Multiplier (GEM),} \emph{NIM. A } {\bf 515} (2003) 249.
	
	\bibitem{datalog} S. Sahu et al., \emph{Design and fabrication of data logger to measure
		the ambient parameters in gas detector R$\&$D}, \emph{JINST} {\bf 12 } (2017) C05006.
	
	\bibitem{NI} \emph{ http://www.ni.com/labview}.
	
	\bibitem{arduino} \emph{https://store.arduino.cc/arduino-due}.
	
	\bibitem{micro} \emph{ Microcontroller Atmel SAM3X8E ARM Cortex-M3 CPU  programming https://www.arduino.cc/reference/en}.
	
	\bibitem{arduino pic} \emph{https://www.digikey.com/en/articles/techzone/2013/aug/increasing-security-
		through-technologies-integrated-into-microcontrollers}.
	
	\bibitem{w5100} \emph{https://www.sparkfun.com/datasheets/DevTools/Arduino/\\W5100{\_}Datasheet{\_}v1{\_}1{\_}6.pdf}.
	
	\bibitem{w5100.1} \emph{https://www.arduino.cc/en/Main/ArduinoEthernetShieldV1}.
	
	
	\bibitem{temp} \emph{https://datasheets.maximintegrated.com/en/ds/DS18B20.pdf.}
	
	\bibitem{temp.1} \emph{https://create.arduino.cc/projecthub/TheGadgetBoy/ds18b20-digital-temperature-
		sensor-and-arduino-9cc806}.
	
	\bibitem{pressure} \emph{http://www.datasheetarchive.com/whats{\_}new/f429e6c2566078620f58cef8f472f9f0.html}.
	
	
	\bibitem{pressure.1} \emph{https://learn.sparkfun.com/tutorials/bmp180-barometr
		ic-pressure-sensor-hookup-}.
	
	\bibitem{humidity} \emph{http://www.micropik.com/PDF/dht11.pdf.}
	
	\bibitem{flow} \emph{http://www.aitech-eng.co.jp/info/honeywell-sensing-airflow-awm2000-datasheet.pdf.}
	
	
\end{thebibliography}
\end{document}